# MECHANISM OF CELL INACTIVATION BY DIFFERENT IONS: DAMAGE INDUCTION PROBABILITIES PER SINGLE TRACKS

Pavel Kundrát[*], Miloš Lokajíček, Hana Hromčíková and Libor Judas
Institute of Physics, Academy of Sciences of the Czech Republic, Na Slovance 2, 182 21 Praha 8, Czech Republic

**Biological effects of proton and light ion irradiations have been studied in detail by analyzing published cell survival data with the help of the probabilistic two-stage radiobiological model. Probabilities of single ion tracks to induce lesions of different severity have been assessed in dependence on their linear energy transfer (LET). The results are presented and their interpretation in terms of $Z_{eff}^2/\beta^2$ (the effective charge over velocity squared) is discussed.**

Biological effects of irradiations by ion beams strongly vary with dose, ion species, and energy (or LET). While the number of tracks contributing to a given dose decreases with LET, the complexity of induced damage is assumed to increase and its reparability to decrease. In this work, the effects of single tracks are studied in terms of probabilities to induce DNA lesions of different severity, aiming to quantify the increase in damage complexity with ion charge and LET.

## METHODS

An effective scheme[1] of the probabilistic two-stage model[2] is used in analyzing survival data for Chinese hamster V79 cells irradiated by protons, $^3$He, C, O, and Ne ions of diverse energies[3-8]. The effective scheme considers only residual lesions not repaired by the cell, i.e. the effects of repair processes are included in the residual damage characteristics. This approach enables to represent differences in the effectiveness of radiations of diverse quality[1,9]; for an explicit treatment of repair processes, reflecting the radiation sensitivity of different cell lines, see Refs.[2,10].

Denoting by $a$ and $b$ the average probabilities per track to induce single-track and combined damage[1,2], the cell survival probability can be expressed as

$$s = \Sigma_k P_k q_k \qquad (1)$$

with the survival probability of cells hit by $k$ tracks[1]

$$q_k = (1-a)^k ((1-b)^k + kb(1-b)^{k-1}), \quad q_0=1, \quad q_1=1-a. \qquad (2)$$

In microbeam irradiation, the corresponding predefined distribution $P_k$ of tracks over cell nuclei has to be considered (and $q_0 \le 1$, $q_1 \le 1-a$, etc. taken to represent the bystander effects). Under broad-beam irradiation, tracks are distributed randomly according to Poisson statistics, $P_k(D)=\exp(-hD)(hD)^k/k!$, where $D$ is applied dose and the mean number $h$ of tracks per nucleus at unit dose[1,2] is proportional to geometric cross section of cell nucleus $\sigma$; in this work $\sigma = 87.8$ μm$^2$ has been used for V79 cells[7].

In Ref.[1], simultaneous analyses of the whole data sets for individual ions were reported, in which damage induction probabilities $a(L)$, $b(L)$ were derived as functions of LET $L$ of a given ion. In this study, damage probabilities have been derived by separate analyses of single survival curves, with the primary aim to benchmark the systematic fits. In addition, the possibility of using the effective charge $Z_{eff}^2/\beta^2$ as a parameter more relevant to biological effects than LET has been examined.

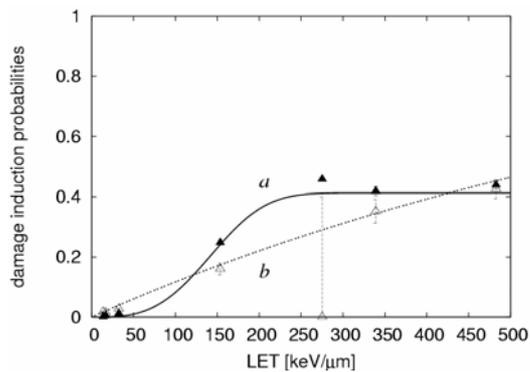

Figure 1: Single-track $a$ (▲) and combined $b$ (Δ) damage probabilities for carbon tracks in V79 cells in dependence on LET. Systematic fits of Ref.[1] shown for comparison (lines).

## RESULTS AND DISCUSSION

In Figure 1 the best-fit values of damage probabilities, derived from single survival curves for carbon ions, are shown and compared to systematic fits to the given data set (Ref.[1]). Similar comparisons have been made for other ions, too (results not presented here). They have verified that although the independent fits suffer from some scatter between the values derived from data from different groups or at different energies, the observed trends with increasing LET agree with the systematic fits. However, for high-energy protons, the independent fits have helped reveal an underestimation of single-track effects in the results presented in Ref.[1]; cf. Ref.[9]. E.g. for 7.7 keV/μm protons, the estimated yields of single-track lethal events are $a.h$=0.04 Gy$^{-1}$ following Ref.[1] vs. $a.h$=0.27 Gy$^{-1}$ from the present results, in a good agreement with the linear-quadratic analysis yielding[3] $\alpha$=0.29 Gy$^{-1}$.

*Corresponding author: Pavel.Kundrat@fzu.cz

of the present work demonstrate that, even at the level of single-track effects, neither LET nor $Z_{eff}^2/\beta^2$ can be used as a single descriptor of radiobiological effects. Further track structure parameters must be considered, such as the effective track diameter and density profile.

**Acknowledgment:** This work was supported by the grant "Modelling of radiobiological mechanism of protons and light ions in cells and tissues" (Czech Science Foundation, GACR 202/05/2728).

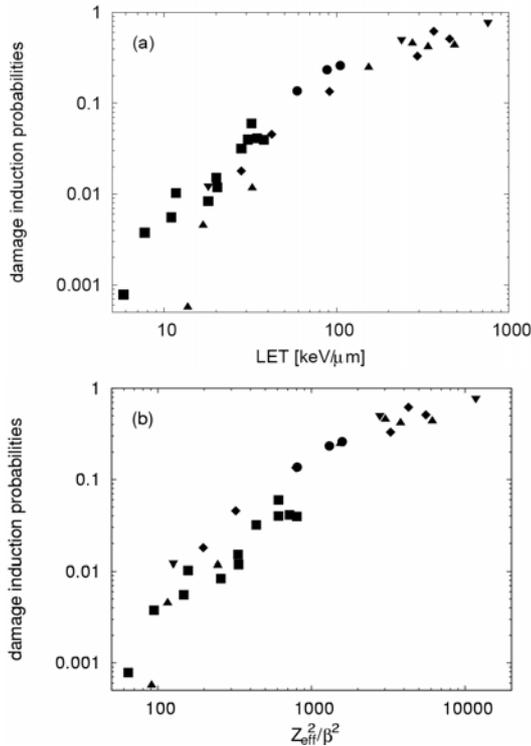

Figure 2: The estimated values of single-track lesion induction as function of LET and $Z_{eff}^2/\beta^2$ for protons (■), $^3$He (●), C (▲), O (▼), and Ne (♦) ions.

Figure 2 shows a common plot of the single-track damage probabilities for all ions in dependence on LET (panel a) or the effective charge $Z_{eff}^2/\beta^2$ (panel b). Damage probabilities in general increase with LET and atomic number $Z$. However, in the LET region below approx. 30 keV/μm, protons tend to be the most effective ions not only in terms of cell survival[3,6], but also at the level of single-track effects. E.g. for LET ~ 15 keV/μm, proton tracks are approx. 10 times more effective than carbon ions in inducing lethal lesions. These differences follow from the track structure: proton tracks are narrower, denser and hence more effective than tracks of heavier ions of the same LET.

The differences in track effectiveness diminish partially when description in terms of effective charge $Z_{eff}^2/\beta^2$ is used (Figure 2b); similar observation was reported in phenomenological analyses of inactivation cross sections of different ions[11]. However, the results